# ISMS ROLE IN THE IMPROVEMENT OF DIGITAL FORENSICS RELATED PROCESS IN SOC's

Masoud Hayeri Khyavi
ICT Security Department
Iran Telecom Research Center
Postcode 1439955471
TEHRAN, IRAN
m.hayery@itrc.ac.ir

## ABSTRACT

Organizations concerned about digital or computer forensics capability which establishes procedures and records to support a prosecution for computer crimes could benefit from implementing an ISO 27001: 2013-compliant (ISMS Information Security Management System). A certified ISMS adds credibility to information gathered in a digital forensics investigation; certification shows that the organization has an outsider which verifies that the correct procedures are in place and being followed. A certified ISMS is a valuable tool either when prosecuting an intruder or when a customer or other stakeholder seeks damages against the organization. SOC (Security Operation Center) as an organization or a security unit which handles a large volume of information requires a management complement, where ISMS would be a good choice. This idea will help finding solutions for problems related to digital forensics for non-cloud and cloud digital forensics, including Problems associated with the absence of standardization amongst different CSPs (Cloud service providers).

## 1. INTRODUCTION

In today's world where cybercrime is increasing constantly, pervasive computing is on the rise and information is becoming the most sought-after commodity making an effective and efficient Information Security architecture and program essential. With this improved technology and infrastructure, ongoing and pro-active computer investigations are now a mandatory component of the Information Security enterprise [1]. The Information Security program of an organization is only as strong as its weakest link. Incidents will occur inevitably, but what matters is finding a connection between the attacker or the source of the attack and a security incident so that management can decide to take the proper action [1].

Information security being systematic is necessary but not sufficient, and here management plays an important role. Through exact management of information security, events procedure can be predicted, controlled and tracked. Inputs and outputs are specified and categorized; the work procedure is specified and responsibilities are clarified. Security Operation Center (SOC) as an organization or a security unit which handles a large volume of information requires a management complement, where ISMS would be a good choice. Importance of this issue is more sensible in Forensics context where preserving proofs and their acceptability is of great importance; additionally, in such situation where information security management system is confirmed, their responsibilities cannot be disclaimed.

ISMS is a management method which has specific standards. This method considers all security aspects from management point of view based on standards which are certified by ISO to distinguish the best way for designing, implementing, running and managing them based on which all assets (tangible and intangible), vulnerabilities, risks, threats and controls would be considered to present a new security comprehensive scheme [2]. Since ISMS is based on a group of security standards in the context of information security, this paper reviews main definitions considering main elements which comprise these standards to propose a solution for improving processes related to forensics in a SOC. It is important to note that standards are reviewed and updated every five years [3].

## 2. RELATED WORKS

A study on Information Systems Audit and Control Association (ISACA) entitled "The computer forensics and cybersecurity governance model" has explained importance of a computer Security incident response team (CSIRT) in every organization and federal agency (Figure 1). Authors have explained that there are three steps required for effective use of the governance model and to understand whether an intersection exists between information security (infosec) and computer forensics or not as follows [4]:

- Step 1: Infosec risk assessment (prevention)
- Step 2: Computer forensic assessment (detection)
- Step 3: Analyzing the intersection between infosec and computer forensics

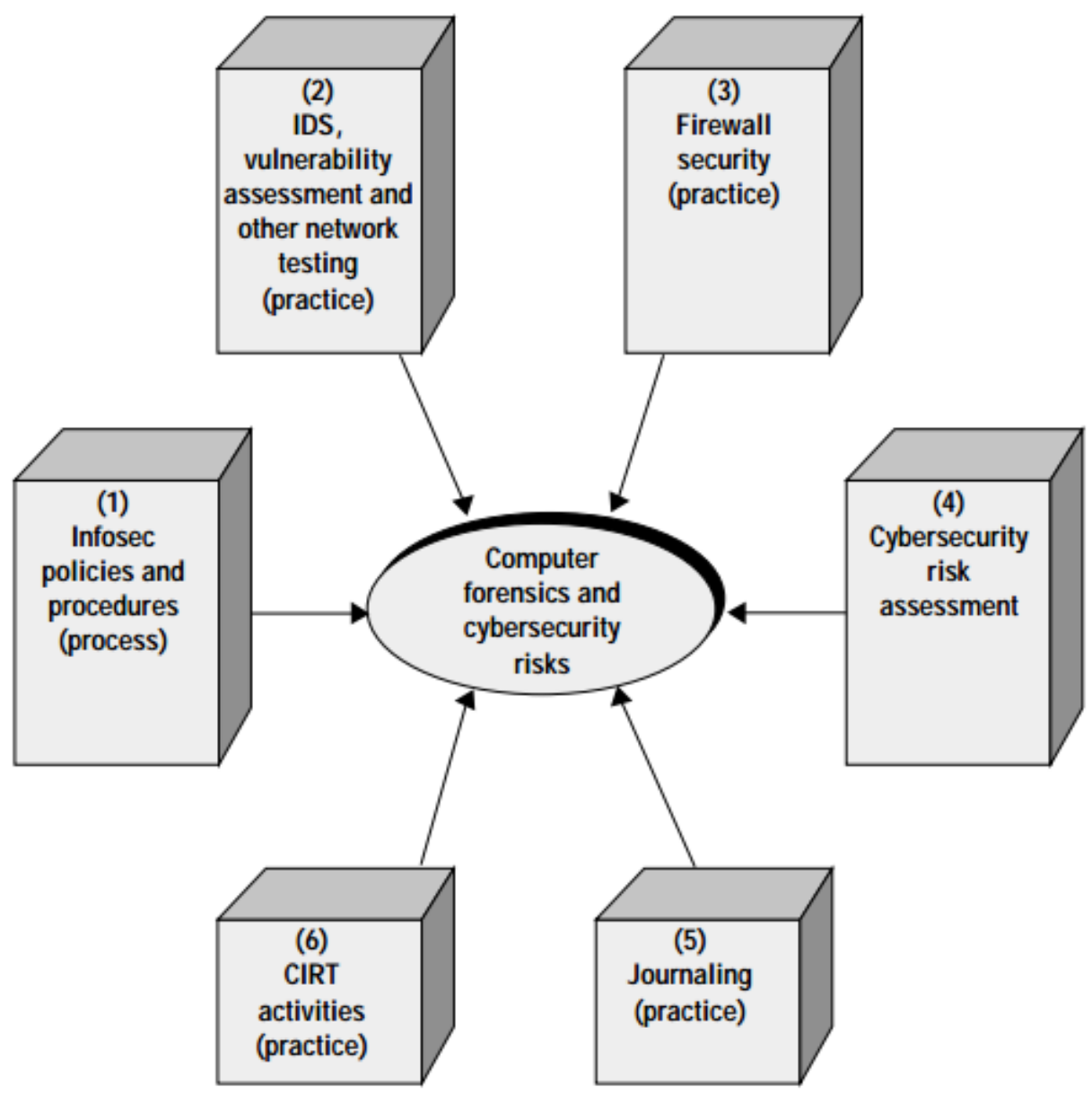

Figure 1. Computer Forensics and Cybersecurity methodology parent process flow diagram [4].

Considering SOC and CSIRTs duties, it can be concluded that CSIRTs can be considered as a specialized SOC [5], [6].

An innovation of this paper is to study efficiency of ISMS in digital Forensics' implementation. It is a complex of our experiences in ICT security department in Forensics, SOC and ISMS implementations.

Before talking about our idea, let's take a look at the main security concepts including ISMS, SOC and forensics.

## 3. SOC AND ITS DIGITAL FORENSICS CHALLENGES

SOC is known as an Organization implemented in a parallel network to keep the organization secure against threats, attacks and unknown vulnerabilities. A SOC is related to people, processes and technologies involved in providing situational awareness through detection, containment and remediation of IT threats [7]. A SOC manages incidents for the enterprise, ensuring they are properly identified, analyzed, communicated, actioned/defended, investigated and reported. The SOC also monitors applications to identify a possible cyber-attack or intrusion (event) and determines if it is a real malicious threat (incident) and if it could have a business impact [7].

Goal of a SOC is to monitor security-related events from enterprise IT assets including IT network, perimeter defense systems such as firewalls, intrusion prevention devices, application servers, databases and user accounts [6].

Effectiveness of a SOC depends on its analytic and forensic capabilities, access to actionable threat intelligence, awareness of the enterprise networks and systems, and internal processes to coordinate responses from organizations across the enterprise. Also, security incident and event management (SIEM) systems have been designed to meet SOC's challenges. Hence, modern enterprise SOCs are structured hierarchically around a SIEM system. Enterprises-scale SIEM systems need significant investment in both hardware and manpower [6]. SIEM systems are an important tool in SOCs—collecting, normalizing and analyzing security events from diverse sources but they must evolve to overcome future scalability issues [6].

Thus, SIEM systems like radars detect objects promptly. Their long-term event retention capability is useful for post hoc forensic analysis as well as investigating and detecting slow and stealthy attacks including advanced persistent threats (APTs) [6].

A typical SIEM system architecture accepts inputs from various security devices and sensors. Connectors receive events, parse them and convert them into a common format [6].

Then SOCs mention three digital Forensics challenges when using SIEM systems [6]:

- Operational: driven primarily by the scale and complexity of the enterprise being monitored and the rate at which events arrive from security devices and sensors, (Lack of qualified forensic analysis personnel available when needed) [6].

- Technical: the scale of events enterprise networks generated from hardware devices, software applications and actions of people in the network and more complex applications and devices trend to generate more events, (Lack of forensic technologies to conduct a thorough and efficient investigation) [6].
- Device Costs: To meet the existing challenges and prevent future challenges that will affect digital forensics aspects and processes in SOC, we recommend implementing an ISMS. SOC can use ISMS capability besides SOC systems like COBIT or ITIL process field

Information Security Management System has an upstream functional Role, but there is a technical Role for SOC and a legal Function for Forensics.

According to the forensics, with its very complicated nature, ISMS methodology should be used in the implementation of forensics procedures in SOC and these points can be used to improve related conditions of processes.

We can manage information security management system, such as a police station, SOC like a SWAT team and Forensics as well as assuming criminal investigation unit.

It should be noted that risk assessment; a part of risk management, is One of the most important domains of information security management which is very important in the procedure of regulatory adoption. For this purpose, this trend is described in the following in three stages:

**Step 1: Digital Forensics Components**

Two approaches can be considered for digital forensics [8]:

1) Proactive Digital Forensics Component: is the ability to collect, trigger an event, preserve and analyze evidence to identify an incident as it occurs, proactively. Also, an automated preliminary report is generated for a later investigation by the reactive component. The evidence that will be gathered in this component is the proactive evidence that is related to a specific event or incident as it occurs. As opposed to the reactive component, the collection phase in this component comes before preservation since no incident has been identified yet [8].
2) Reactive Digital Forensics Component: is the traditional (or post-mortem) approach of investigating a digital crime after an incident has occurred. This involves identifying, preserving, collecting, analyzing and generating the final report. Two types of evidence are gathered under this component: active and reactive. Active evidence refers to collecting all live (volatile) evidences that exist after an incident. An example of such evidence is the processes which run in memory. The other type, reactive evidence, refers to collecting all the remaining static evidences, such as image of a hard drive [8].

There are four key aspects in digital evidence handling also: auditability, justifiability and either repeatability or reproducibility depending on particular circumstances.

It is highly desirable that the investigation of an incident should be planned in advance, but there are circumstances (e.g., when an incident is developing and being responded to in real-time) where full planning is not be possible. In such situations, the team should brief on initial strategies and tactics for the investigation and should be allowed to develop new strategies and tactics in response to prevailing conditions.

As the incident develops, information about it should be shared amongst the team as quickly as possible to ensure that proper actions are taken efficiently regarding the need for justification.

**Step 2: Information Security Management System and Digital Forensics**

Information Security can be defined as the process of protecting information and information assets from a wide range of threats in order to ensure business continuity, minimize business damage, maximize return on investments and business opportunities by preserving confidentiality, integrity and availability of information [1].

Aside from its roots in forensic sciences, computer or digital forensics are also considered as sub-disciplines of the information security field. The idea behind computer forensics which is well established and simple is that electronic evidence is discoverable [9]. Every data, file, or document stored in computer (on a permanent or temporary basis or transiting by a computer component) can be inspected using a unique string, or ‘‘digital fingerprint,’’ which can even represent an entire computer’s hard drive. Furthermore, computer forensics inspections can recover erased data, decrypt encrypted information and derive hidden information or evidence from a suspicious computer [9].

Forensics is using science and technology to investigate and establish facts in criminal and civil courts of law; goal of any forensic investigation is to prosecute the criminal or offender successfully, determine the cause of an event and determine the one who was responsible [1].

It should be considered that Digital Forensics readiness will concern itself with incident anticipation - instead of incident response - and enabling the business to use digital evidence. Information security will concern itself by ensuring the business utility of information and information assets is maintained excluding the requirement for digital evidence [1]. We have their overlap. For example, there should be clear segregation of duties between the digital forensics and information security teams.

In table 1, DF is used for Digital Forensics and IS for Information Security:

Table 1- Comparison of Information Security and Digital forensics [1]

| | Information Security | Digital Forensics |
|---|---|---|
| Pro-active | **Purpose**:<br>• Prevent damage to information and information resource by applying the most effective and efficient controls for the identified threats by the organization | **Purpose: (two fold)**<br>• Ensure that all process, procedures, technologies and appropriate legal admissible evidence is in place to enable a successful investigation, with minimal disruption of business activities<br>• Use DF technology to determine the 'holes' in the security posture of the organization |
| | **How:**<br>• IS policies, e.g. incident recognition, Implement IS procedures and mechanisms<br>• Determine legal requirements<br>• IS awareness and training | **How:**<br>• DF policies, e.g. evidence preservation<br>• DF readiness, e.g. Prevent anonymous activities, secure storage of logs, hashing etc.<br>• Determine legal requirements<br>• DF awareness and training |
| Re-active | Purpose:<br>• Ensure that the damage that has occurred from a breach is minimized and prevent further damages | **Purpose**:<br>• To investigate an event in a way that the evidence gathered can be used to determine the root-cause of an event and successful prosecution of the perpetrator |
| | **How**:<br>• Incident response plan(IRP)<br>• Disaster Recovery Plan(DRP)<br>• Business Continuity Plan(BCP) | **How**:<br>• Incident response plan(IRP)<br>• Disaster recovery plan(DRP)<br>• Business Continuity plan(BCP) |

In "Overview of Digital Forensics," ISACA recommends seven considerations for inclusion in the information systems' life cycle to effectively handle incidents in case they were to suffer a cyber-attack [10]:

1) Perform regular system backups and maintain previous backups for a specific period.

2) Enable auditing on workstations, servers and network devices.

3) Forward audit records to secure centralized log servers

4) Configure mission-critical applications to perform auditing and record all authentication attempts.

5) Maintain a database of file hashes for the files of common operating system and application deployments, and use file integrity checking software on particularly important assets.

6) Maintain records (e.g., baselines) of network and system configurations.

7) Establish data retention policies that support historical reviews of system and network activity, comply with requests or requirements to preserve data that are related to ongoing litigation and investigations and destroy data that are no longer needed.

An understanding of digital forensics is essential to ensure protection of an organization in the event of a security breach that involves loss of confidential information. It is also a key element in implementing and maintaining an effective ISMS as specified by Control A.13.2.3 of the ISO 27001 standard which requires that, in the event of a security incident, any evidence presented in a criminal or civil action against an individual or company must fully conform to all relevant legislation [11].

There are four main forensic-related controls that an ISO 27001:2013–compliant organization must address [12]:

- A.16.1.1— Responsibilities and procedures

  Management responsibilities and procedures should be established to ensure a quick, effective and ordered response to information security incidents [11].

- A.16.1.6— learning from information security incidents

  There should be mechanisms in place to enable the types, volumes and costs of security incidents to be quantified and monitored [11].

- A.16.1.7— Collection of evidence

  Where follow-up action against a person or organization after a security incident involves legal action (either civil of criminal); evidence should be collected, retained and presented to conform to the rules of evidence laid down in the relevant jurisdiction(s) [11]. Also, a control about Back-up confronted with anti-forensics actions can be used.

- A.12.3.1— Information back-up

  Using back-up for recovering data cleared through anti-forensic techniques used by targeted attackers via Volume Shadow Copy and Restore Point analysis [11], [13].

And following Auxiliary control:

A.16.1.4—Assessing and deciding about information security events

In order to avoid interference of functions, the following items can be used:

- A.6.1.2 Segregation of duties

  And for Training and awareness as shown in table 1, we have the following control:

- A.7.2.2 Information security awareness, education and training

  Digital forensics awareness training is linked to Information security awareness training, for example with first incident response training [1].

In addition, while this requirement is quite obvious, it is crucial for success of the legal process, that the digital evidence is collected as accurately and reliably as possible. The best practice as defined in clause 16.1.7 of the ISO 27002: 2013(not a management standard, only best practice, cannot be accredited). This Standard recommends procedures for a forensic collection of digital evidence [14]. Another helpful standards in this regard are ISO 27041:2015 and ISO 27043:2015. All such plans are major contributors to ensure conformance to Clause 7.3 of the ISO 27001 Standard on preventative action which is, of course, essential to the maintenance of ISMS continual process improvement [14].

There is digital forensics' related Standards, ISO 27037:2012 [15] and ISO 27042:2015. In ISO 27037, the complete digital evidence handling process includes other activities (i.e., presentation, disposal, etc.). The scope of this International Standard is only related to the initial handling process which consists of identification, collection, acquisition and preservation of potential digital evidence and so these four processes are considered here.

**Step 3: Improving Digital Forensics Procedures in SOC Implemented with ISMS**

There question is: how can ISMS help and assist in SOC's Forensics Process?

The most challenging and important part of Digital Forensic Investigation (DFI) is data examination [16]. Main Standard about ISMS is ISO 27001, and this standard provides a framework for implementing a certifiable ISMS, to support a digital forensic investigation [17]. With forensic procedures in place, everyone involved in a forensic investigation will understand their responsibilities [12].

The deployment of an ISMS can address many other business risks which an organization may run, giving rise to additional benefits. The ISMS team can use statistical information on types, costs and frequency of incidents to establish priorities for improvement processes [12].

Overall, implementation of ISO 27001 will require an organization to plan for a more efficient incident response or management. Controls will be identified to reduce the incidence of identifiable incidents and to detect the incidents when they occur. Incident investigation procedures will be developed and the required personnel will be identified and trained. Legal obligations will be identified and included in the maintenance of records and the incident handling procedures [12]. Moreover, by using ISO 27001 controls about Back-up, the Anti-forensic actions can be thwarted via Recovering data cleared through anti-forensic techniques used by targeted attackers.

From a digital investigation perspective, anti-forensics can do the following [8]:

- Prevent evidence collection.
- Increase the investigation time.
- Provide misleading evidence that can jeopardize the whole investigation and subsequent lawsuit.
- Prevent detection of digital crime.

In short, ISO 27001 provides the pre-incident planning that makes a forensic investigation possible. It provides a management environment that is favorable for investigations [12].

In case, the responsibilities are identified correctly and completely by the standard; it would play an important role in the prevention of crime in Cyberspace plays. Also, part of the new standard ISO 27037 can be used.

## 4. Mapping

In this stage, we have a mapping to model the ISMS recommendations with digital forensics indicator. To provide simplicity and sharpness of the model, check and act steps are merged into one section.

How this mapping has outlined the steps, will be described later:

- **Plan** included:
  - Risk Management
  - Controls
  - Audit
- **Do** included:
  - Network security teams' activities
- **Check** and **Act** included
  - Audit trail
  - Forensics investing
    - Proactive digital components
    - Reactive digital components
  - Incident management

Figure 2 can be used employing ISMS controls:

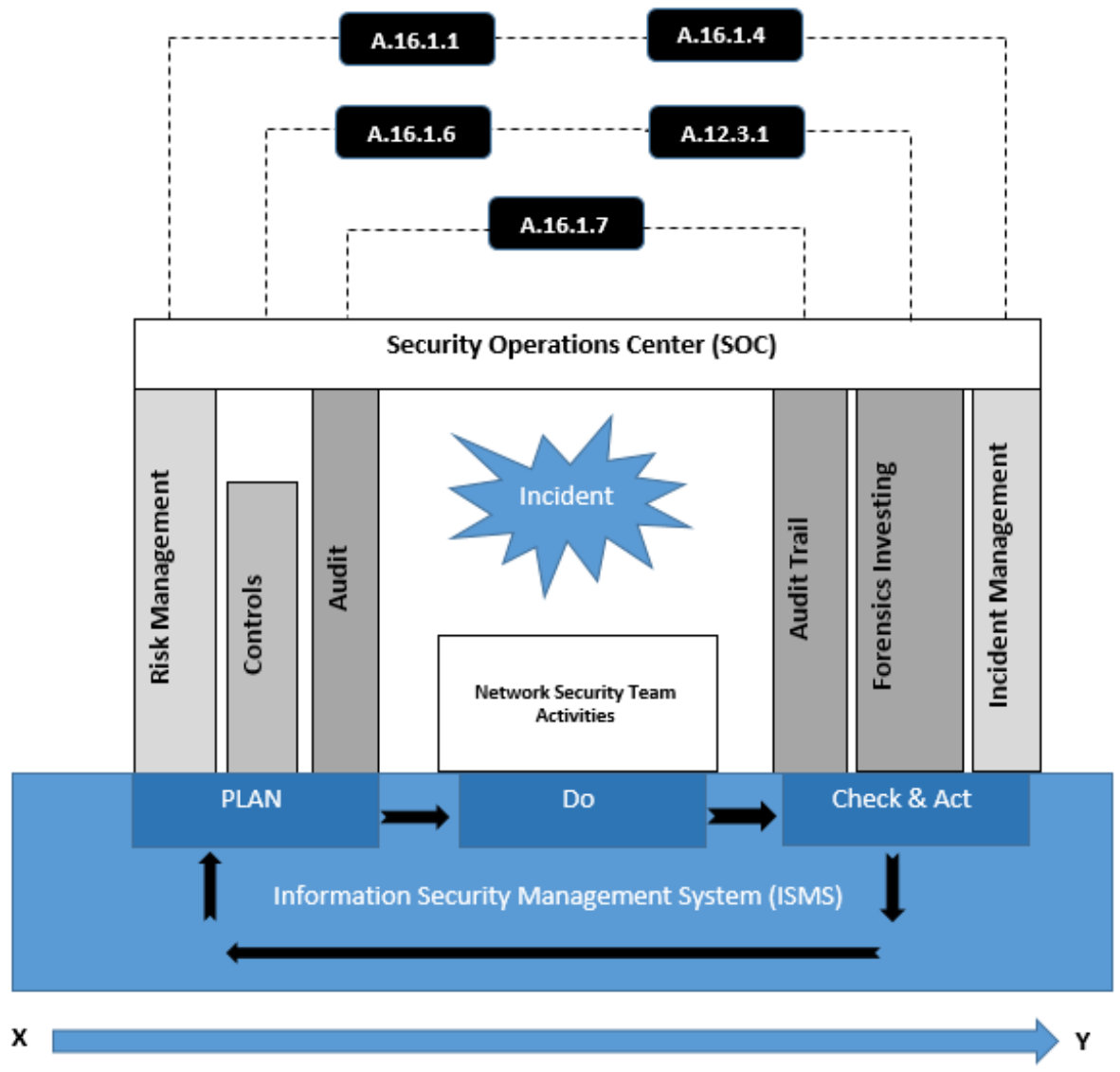

Figure 2. Important components in preventing and dealing with crime in the context of ISMS

On the other side, the following figure (figure 3) illustrates a special area that can be considered as a result of two other controls (A.6.1.2 and A.7.2.2).

These important controls are related to handling, and response team in which human factors (Accuracy, Knowledge and Response) are involved.

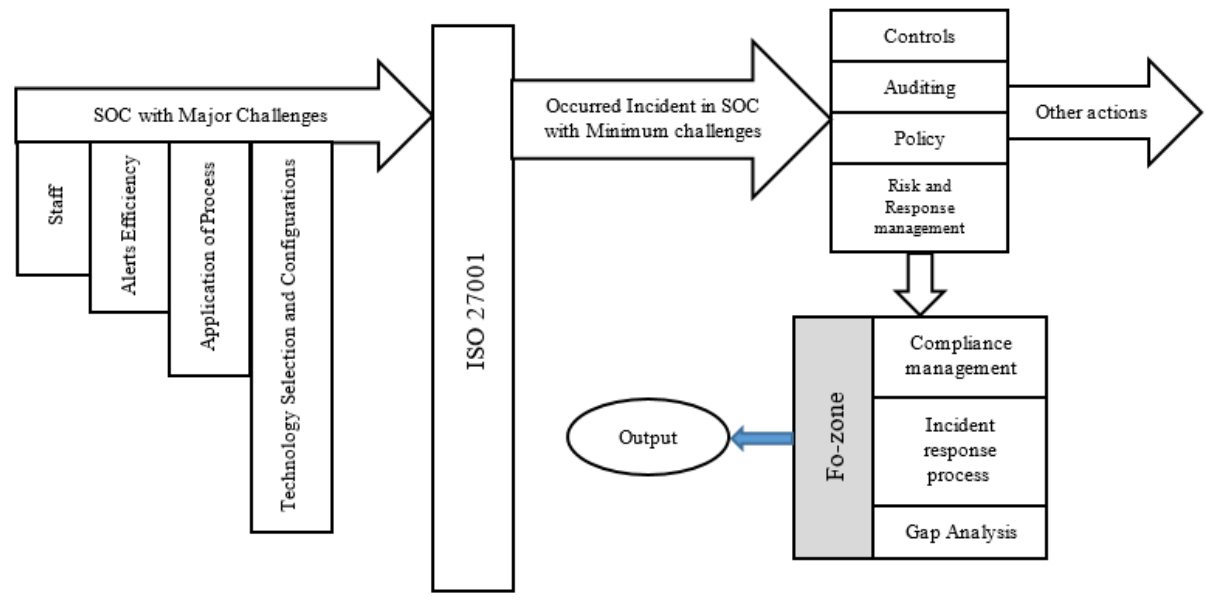

Figure 3. Forensic' texture in SOC

In Figure 4, an area is shown in gray that explains following cases in computer or digital forensics and it is called Fo-Zone [1](Forensic Handling Zone – Term recommendation). Four ISMS controls can be used to reduce the area on the Forensics (Fo-Zone) and another control can be used to facilitate handling of Forensics.

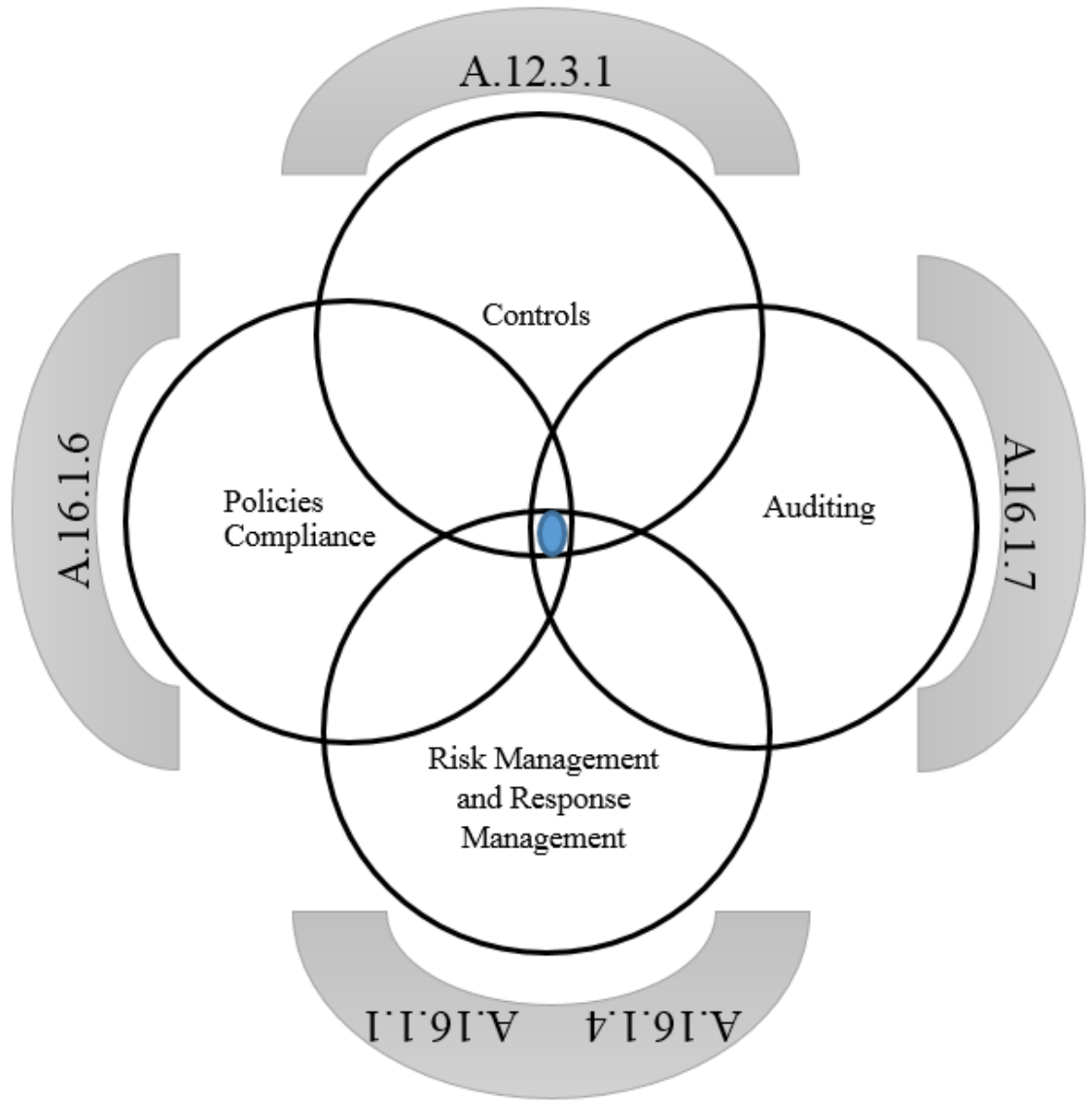

Figure 4. Fo-Zone and the Main Controls

Fo-Zone plays a critical role in Compliance management, incident response process and gap analysis. Also, four key dimensions can be seen: Technical, Human, Organizational and Regulatory.

[1] Forensics Handling Zone

These dimensions and their aspects can be used in the Comprehensive roadmap (research agenda) to fight against cybercrime and cyber terrorism (CAMINO) in which, forensics plays an important [18].

The main contribution of this idea is to use ISMS capability in SOC for incident responding from Forensics point of view in shortest time with highest accuracy. On the other hand, compliance management in SOC can be improved.

In each crime, time and accuracy for saving the evidence which are reliable are the most critical factors. If you pay more attention to the picture, it would be clear that by ISMS specialties, the best platform for SOC can be created. This image has a conceptual meaning. Using standards such as ISO 27001:2013 makes guidelines and procedures related to SOC's forensics more structured and validated.

An efficient network security work starts way before incident. It starts with training and preparation and analysis of risk and the reaction of the team starts with a ansident. Suppose that controls in the image are LED lamps which turn on by an incident after which operation will start. Other controls can be added from more related security standards in the control block.

## 5. Conclusion

Building an enterprise security operations center can be an effective path to reduce security vulnerabilities. An enterprise or another type of SOC encompasses people, processes and technologies that handle information technology threat monitoring, forensic investigation, incident management and security reporting [19].

ISO 27001:2013 and ISO 27002:2013 provide the pre-incident planning that makes a forensic investigation possible. Controls will be identified to reduce the frequency of incidents and to detect incidents when they occur. Incident investigation procedures will be developed and the required personnel will be identified and trained. It provides a management environment that is favorable to investigations [12].

But as it was seen, technical and non-technical challenges are covered in forensics procedures implemented in SOC, and it could be done with an ISMS. It is known that it is expensive to safeguard sensor and computer networks in general, because there are diversities of sensor nodes and multiple resources to be protected from intruders. Moreover, suitable tools for conducting network forensic investigations are generally expensive at the moment [20].

Besides, existing tools are resource intensive regarding replacement cost, installation and maintenance costs, processing time, disk space and skill required to convert audit logs into admissible evidence in courts of law. Therefore, the development of affordable tools that are available to the research community is a classical problem in forensic investigations of sensor networks [20].

In today's cloud computing environment, organizations which want to reduce costs without compromising information security are looking at ISO 27001 certification as a promising means for providing knowledge about their IT security [10] [25]. ISMS can foster efficient security (digital forensics) cost management in a SOC, compliance with laws and regulations, and a comfortable level of interoperability due to a common set of guidelines followed by the partner organization [21]. This idea will help finding solutions for problems related to digital forensics for non-cloud and cloud digital forensics, including Problems associated with the absence of standardization amongst different CSPs [3].

Another contribution of this idea is to have a hybrid approach. In many cases related to security jobs, especially in incident responding and Forensics, work interferences are faced which have a negative impact on incident responding and incident handling. Generally speaking, users of ISO 27001 prefer to consider all controls (in projects related to certification).

According to Figure 4, it can be used as a reference which can be discussed and the connection between items and surely new ideas can be created.

## References

[1] C P Grobler; C P Louwrens, "Digital Forensic Readiness as a Component of Information Security Best Practice," in *New Approaches for Security, Privacy and Trust in Complex Environments*, Sandton, 2007.

[2] Masoud Hayeri Khyavi; Mina Rahimi, "Conceptual model for security in next generation network," in *30th International Conference on Advanced Information Networking and Applications Workshops (WAINA)*, Crans-Montana, 2016.

[3] Reza Montasari ; Richard Hill , "Next-Generation Digital Forensics: Challenges and Future Paradigms,"

in *12th International Conference on Global Security, Safety and Sustainability (ICGS3)*, London, 2019.

[4] K. C. Brancik, "The Computer Forensics and Cybersecurity Governance Model," *INFORMATION SYSTEMS CONTROL JOURNAL,* pp. 41-47, 2003.

[5] Pierre Jacobs et al, "Classification of Security Operation Centers," in *Information Security for South Africa (ISSA)*, Johannesburg, 2013.

[6] Sandeep Bhatt et al, "The Operational Role of Security Information and Event Management Systems," *IEEE Securit & Privacy Magazine,* pp. 35-41, 2014.

[7] 2016. [Online]. Available: http://www.webeyesystems.com.

[8] Soltan Alharbi et al, "The Proactive and Reactive Digital Forensics Investigation Process: A Systematic Literature Review," in *International Conference on Information Security and Assurance*, Brno, 2011.

[9] Yun Wang et al, "Foundations of computer forensics: A technology for the fight against computer crime," *Computer Law & Security Review,* pp. 119-127, 2005.

[10] ISACA, "ISACA," [Online]. Available: http://www.isaca.org.

[11] "castleforce," [Online]. Available: http://www.castleforce.co.uk.

[12] ISO/IEC 27001, Information technology — Security techniques — Information security management systems, ISO, 2013.

[13] SANS, "Advanced Digital Forensics and Incident Response," SANS, 2015.

[14] ISO/IEC 27002, Information technology – Security techniques – Code of practice for information security controls., ISO, 2013.

[15] ISO/IEC 27037, Information technology – Security techniques - Guidelines for identification, collection, acquisition and preservation of digital evidence, ISO, 2012.

[16] Yazid Haruna Shayau; Aziah Asmawi; Siti Nurulain Mohd Rum; Noor Afiza Mohd Ariffin , "Digital forensics investigation reduction model (DIFReM) framework for Windows 10 OS," in *9th International Conference on System Engineering and Technology (ICSET)*, Shah Alam, 2019.

[17] Stavros Simou; Christos Kalloniatis; Stefanos Gritzalis; Vasilios Katos , "A framework for designing cloud forensic-enabled services (CFeS)," *A framework for designing cloud forensic-enabled services (CFeS),* p. 403–430, 2019.

[18] Michal Choras et al, "Comprehensive roadmap (research agenda) for fight against cybercrime and cyber terrorism," in *10th International Conference on Availability, Reliability and Security*, Toulouse, 2015.

[19] IBM, "IBM," 2013. [Online]. Available: https://www.itdigitalsecurity.es.

[20] Joshua Ojo Nehinbe; Peter Damuut , " Security issues in Sensor Networks and gathering admissible evidence in Network Forensics," in *UKSim 5th European Symposium on Computer Modeling and Simulation*, Madrid, 2011.

[21] ISACA, "Planning for and Implementing ISO 27001," 1 July 2011. [Online]. Available: https://www.isaca.org.

[22] ISO/IEC, Information technology – Security techniques – Code of practice for information security controls. The latest version of the code of practice for information security controls, ISO, 27002: 2013.